\documentclass[11pt, a4paper, twocolumn, copyright]{google}

\usepackage[numbers,sort&compress]{natbib}
\usepackage{graphicx}
\usepackage{siunitx}
\usepackage[utf8]{inputenc}
\usepackage[T1]{fontenc}
 \usepackage{url}
 
\usepackage{placeins}
\usepackage{amssymb}

\title{Towards a future space-based, highly scalable AI infrastructure system design}

\correspondingauthor{jessicavbloom@google.com}

\renewcommand{\today}

\author[*,1]{Blaise Ag\"uera y Arcas}
\author[*,1]{Travis Beals}
\author[*,1]{Maria Biggs}
\author[*,1]{Jessica V. Bloom}
\author[*,1]{Thomas Fischbacher}
\author[*,1]{Konstantin Gromov}
\author[*,1]{Urs K\"oster}
\author[*,1]{Rishiraj Pravahan}
\author[1]{James Manyika}

\affil[*]{Equal contributions}
\affil[1]{\thepa{}{}}

\begin{abstract}
    If AI is a foundational general-purpose technology, we should
    anticipate that demand for AI compute---and energy---will continue
    to grow. The Sun is by far the largest energy source in our solar
    system, and thus it warrants consideration how future AI
    infrastructure could most efficiently tap into that power. This
    work explores a scalable compute system for machine learning in
    space, using fleets of satellites equipped with solar arrays,
    inter-satellite links using free-space optics, and Google tensor
    processing unit (TPU) accelerator chips. To facilitate
    high-bandwidth, low-latency inter-satellite communication, the
    satellites would be flown in close proximity. We illustrate the
    basic approach to formation flight via an 81-satellite cluster of 1
    km radius, and describe an approach for using high-precision
    ML-based models to control large-scale constellations. Trillium
    TPUs are radiation tested. They survive a total ionizing dose
    equivalent to a 5 year mission life without permanent failures,
    and are characterized for bit-flip errors.  Launch costs are a
    critical part of overall system cost; a learning curve analysis
    suggests launch to low-Earth orbit (LEO) may reach
    $\lesssim$\$200/kg by the mid-2030s.
\end{abstract}

\begin{document}

\maketitle

\section{Introduction}

\subsection{Motivation for ML in space}

The development of the Transformer model \cite{vaswani2017attention}
and the subsequent rise of generative, multimodal AI have led to an
explosion of demand for compute capacity. Although dramatic gains have
been made in efficiency (e.g. Gemini query energy consumption was
reduced $33\times$ over a one year period
\cite{elsworth2025measuring}), AI-based products and services have
grown even faster, leading to a rapid increase in data center energy
demand. To that end, Google is invested in advancing new forms of
power generation (e.g. \cite{Google2024KairosNuclear,
  CFS2025GoogleFusion, Google2023FervoGeothermal}). However, given AI
appears to be a foundational general-purpose technology
\cite{Baily2025generative}---akin to electricity or the steam engine---we
should anticipate that its use will continue to broaden across all
aspects of human endeavor from powering the economy, to helping to
tackle some of humanity's greatest challenges
\cite{gohr2025artificial, mcduff2025towards,jumper2021highly}, and it
can be expected that AI computational needs will grow, as will the
energy required to run it.

In this paper and the research ``moonshot'' it proposes, we look to the
future and work back from that. The Sun is by far the largest source
of power in our solar system: with an output of $3.86 \times
10^{26}$\,W, the Sun emits more than 100 trillion times humanity's
total electricity production. At some point in the future, the best
way to power AI will likely thus be to more directly tap into that
enormous source of energy.  Space-based solar power has long been
proposed---first in a 1941 Asimov short story, 'Reason'. A potentially
feasible technical architecture was proposed in
\cite{glaser1968power}, and recent technical discussions include
\cite{Rodgers2024SpaceBased, kruft2023techno}. It has many of the
attractive qualities of terrestrial solar power, with the additional
advantages that solar panels in certain orbits are exposed to nearly
continuous sunshine, and receive up to $8\times$ more solar energy per
year than a panel located on Earth at mid-latitude
\cite{Sengupta2018NSRDB}. However, getting the generated power back to
Earth has been a major challenge for such proposals.

Instead of transmitting power to Earth from space, we propose a future
that includes space-based ML ``data centers'' consisting of many
solar-powered satellites networked via free-space optical
inter-satellite links. While there are a number of challenges that
would need to be addressed to realize this ``moonshot,'' in the long run
it may be the most scalable solution, with the additional benefit of
minimizing the impact on terrestrial resources such as land and water.

\subsection{System design overview}

Working backward from an eventual future in which the majority of AI
computation happens in space, we identify as an intermediate milestone
showing that a space-based system could achieve performance roughly
comparable to a terrestrial datacenter. This research initiative is
focused on addressing several of the major ingredients required: power
generation, high-bandwidth, low-latency communication between chips,
radiation-tolerant compute, a thermal management system, and a data
link to ground stations. We aim to be as mass-efficient as possible,
maximizing compute per kilogram to minimize launch costs, while
attending to practical considerations for satellite design, including
launch vehicle compatibility, avoidance of space debris, and
structural feasibility.

To meet these requirements, we propose working towards a future where
we would host the Google tensor processing unit (TPU) accelerator
chips on a constellation of solar-powered satellites, with size and
number of TPUs per satellite determined by both economic and
engineering considerations. We envision launching the satellites into
dawn-dusk, sun-synchronous low-Earth orbit (LEO) to enable near-continuous
power generation with lower latency and launch costs than higher orbits.
To enable ultra-high bandwidth, low-latency data transfer
between satellites, they will fly close together and communicate via
free-space optics inter-satellite links (FSO ISLs). An ML-enhanced flight
control model enables the satellites to maintain close flight
proximity while avoiding collisions. Eventually, optical links will
also be needed for high-bandwidth communication with the ground, but
for a pilot project, radio suffices, avoiding the challenges of
overcoming atmospheric interference. Maintaining a dawn-dusk orbit
will increase latency to some ground locations, but is advantageous
for maximizing power. Cooling would be achieved through a thermal
system of heat pipes and radiators while operating at nominal
temperatures.

Proposals exist for ``monolithic'' data centers in space where
individual spacecraft significantly exceed the size of any current or
planned launch vehicle \cite{BBC2025DC,Starcloud2025AIinSpace}. While
such design concepts reduce the need for high-performance
inter-satellite links, they involve new challenges: such structures
would have to be assembled in space by humans or robots; collision
avoidance would be more cumbersome; and structural requirements would
add mass and complexity. Our proposed approach would instead rely on
arrays of smaller satellites. This more modular design would provide
ample opportunity to scale to the terawatts of compute capacity that
could fit within the dawn-dusk sun-synchronous low-earth orbital band.

To assess the viability of this concept, this work focuses on several
key technological challenges: the required inter-satellite
communication bandwidth, the dynamics and control of large,
tightly-clustered satellite formations, the radiation tolerance of
TPUs, and economic feasibility given expected future launch
costs. Other significant challenges such as on-orbit reliability and
repair, high-bandwidth ground communications, and thermal management
are also discussed in this paper. Our ongoing research towards
achieving this moonshot involves refining designs and reducing risks
through further analysis, ground-based testing, and in-orbit prototype
missions---similar to how Google has approached other ambitious
research initiatives.

\section{Results}

\subsection{Inter-satellite links}

The networking requirements of large-scale terrestrial machine
learning (ML) clusters far exceed the capabilities of current
inter-satellite link (ISL) technology. Google's TPU supercomputers,
for instance, utilize a two-tiered networking architecture. A
high-speed data center network provides pod-level connectivity
\cite{singh2015jupiter}, while a custom, low-latency optical
Inter-Chip Interconnect (ICI) with throughputs on the order of
hundreds of gigabits per second per chip facilitates the
tightly-coupled communication required for large-scale training
workloads \cite{jouppi2023tpu}. In contrast, commercially available
optical ISLs offer data rates in the range of 1--100\,Gbps.

\begin{figure}[hbt!]
  \centering

 \includegraphics[width=\columnwidth]{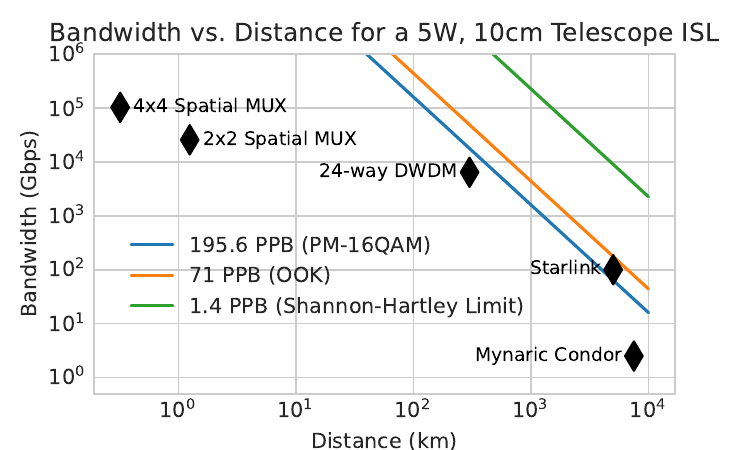}

  \caption{Existing OISL device specifications vs. proposed
    design. Lines illustrate the $1/d^2$ relationship between distance
    and achievable bandwidth for three modulation schemes with
    different photons-per-bit (PPB) requirements. Commercial systems
    operate at long ranges, while our proposed system targets much
    shorter ranges to achieve higher data rates. 24-way dense
    wavelength-division multiplexing (DWDM) can be achieved up to
    about 300km distance with a 10cm aperture size. Fitting $2 \times
    2$ and $4 \times 4$ spatially multiplexed beams
    into the same total aperture requires distances of 1.25km and 0.32km
    (limited by the imaging resolution of each
    sub-aperture rather than by received power). Modulation
    schemes shown: Quadrature-Amplitude modulation with 16 symbols
    (PM-16QAM), on-off keying (OOK), and the Shannon-Hartley limit of
    channel capacity.}
  \label{fig:isl}
\end{figure}

Our analysis shows that the required aggregate bandwidth per link, on
the order of 10\,Tbps, is achievable by using Commercial Off-The-Shelf
(COTS) Dense Wavelength Division Multiplexing (DWDM) transceiver
technology operating in the infrared, similar to that used in terrestrial data centers. The
primary challenge is that such equipment requires significantly higher
received optical power levels, on the order of hundreds of microwatts,
compared to the $\sim$\SI{1}{\micro\watt} levels typical for
traditional long-range ISLs. These power levels can be achieved by
drastically reducing the inter-satellite distance. Since for distances
larger than the Fresnel limit, received power scales with the inverse
square of the distance due to beam divergence, flying the satellites
in close formation (hundreds of kilometers, or less) provides ample
power to close the link budget for high bandwidth COTS transceivers,
as illustrated in Figure \ref{fig:isl}.

As the distance becomes very short (e.g., $\sim$5km for a 10 cm
telescope), spatial multiplexing emerges as a new opportunity for
further scaling. The smaller beam spot size at shorter distances
allows multiple independent beams to be established between
transceiver arrays on different satellites, each carrying a separate
DWDM datastream.  For example, as illustrated with three examples on
the left of Figure \ref{fig:isl}, a single 10 cm total aperture can be
populated with a $2 \times 2$ array of independent 5 cm optical
systems at a link distance of 1.25 km, or a $4 \times 4$ array of 2.5
cm systems at 0.32 km, scaling the total bandwidth with the number of
parallel links.  This enables further scaling of bandwidth inversely
with distance, analogous to achieving high aggregate bandwidth through
parallel spatial streams in Google's Palomar Optical Circuit Switch
\cite{urata2022mission}. In this spatially multiplexed regime, the
use of coherent detection provides inherent rejection of cross-talk
from adjacent beams.

A bench-scale demonstrator using off-the-shelf components successfully
achieved 800\,Gbps unidirectional (1.6\,Tbps bidirectional)
transmission across a short free-space path, validating the potential
of this approach.

\subsection{Orbital dynamics}

\begin{figure*}[t]
  \centering
  \includegraphics[width=\textwidth]{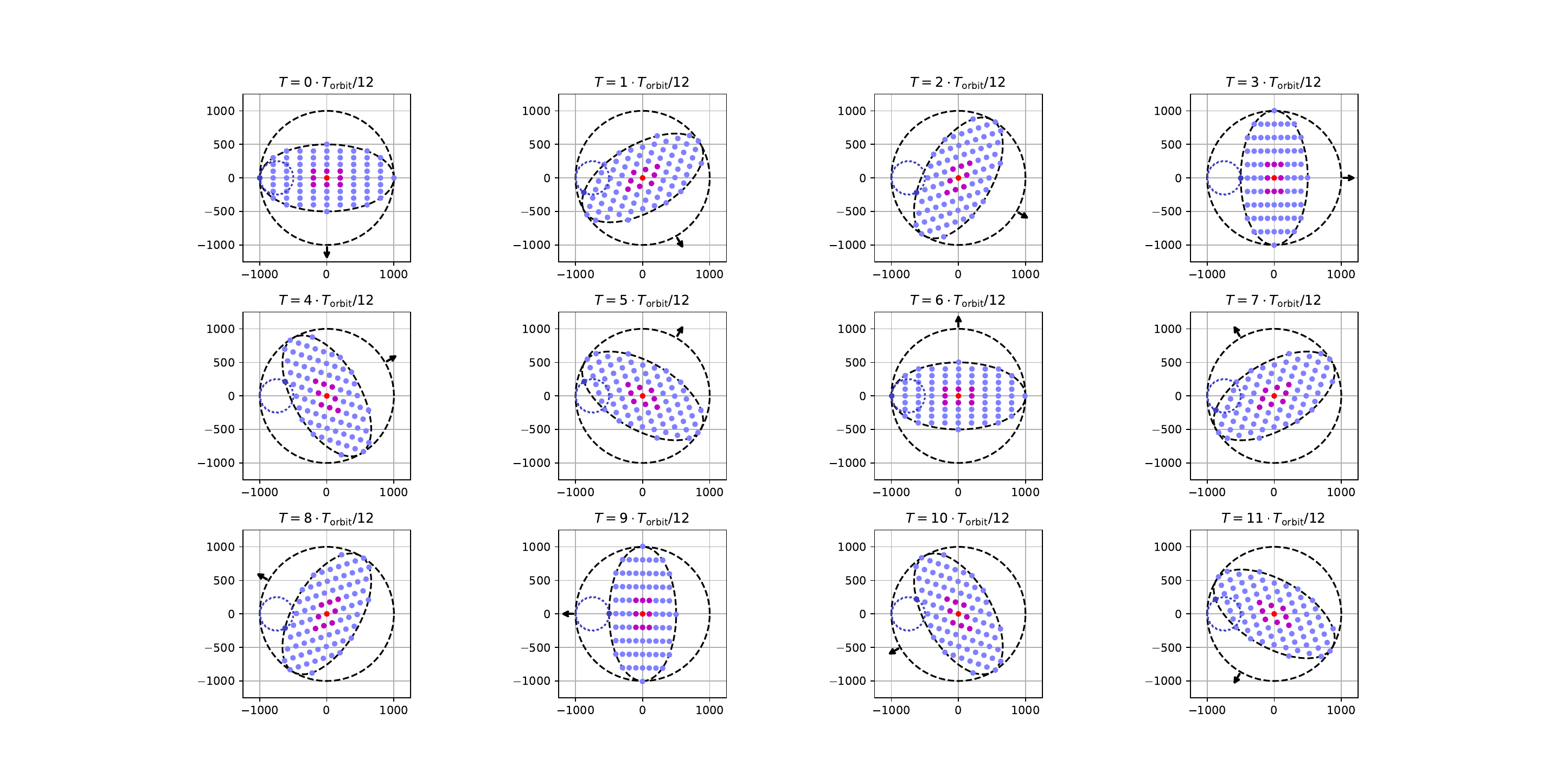}

  \caption{Evolution of a free-fall (i.e. ``no thrust'') constellation
    subject to Earth's gravitational attraction plus J2-term (due to
    Earth's oblateness) over the course of one orbit, shown at time
    intervals of $1/12$ of a full orbit in a non-rotating coordinate
    system. Positions are relative to the central reference satellite
    S0 (red). Horizontal axis is aligned with the negative in-track
    direction of S0 at t=0, vertical direction correspondingly is
    ``towards zenith at t=0.'' Short arrows indicate the ``towards center
    of Earth'' direction. Magenta: nearest neighbors (8-neighborhood)
    of the central satellite S0. Dark blue: the ``maximally-distant in
    in-flight direction at $t=0$'' satellite S1. Dark blue dashed: S1's
    cluster-center-relative positions over the course of one
    orbit. All distances in meters.}
  \label{fig:cluster_orbit}
\end{figure*}

\begin{figure}[htbp]
  \centering

\includegraphics[width=\columnwidth]{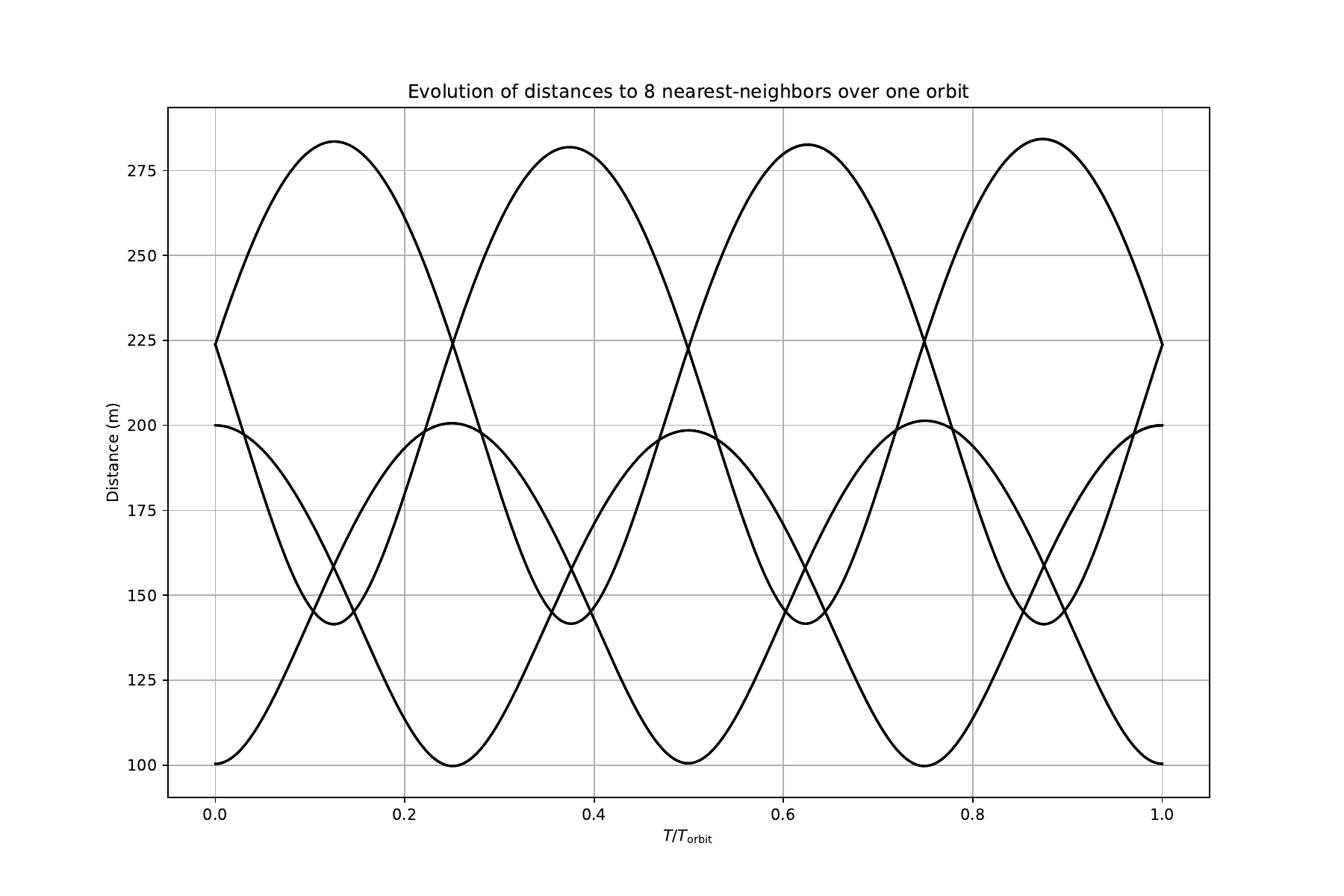}

  \caption{Evolution of the distance between central "reference"
    satellite S0 and its (direct and diagonal) nearest neighbors over
    the course of one orbit under the combined effect of Newtonian
    Gravity and Earth's J2-term.}
  \label{fig:nndist8}
\end{figure}

Our proposed system design, at scale, will be significantly larger,
and entail much closer formation flight (due to inter-satellite
communications requirements), than any previous or current satellite
constellations. We considered a set of constraints including:
maintaining a stable set of nearest neighbors, minimizing latency and
maximizing solar exposure and ISL performance. Based on these
constraints, Figure \ref{fig:cluster_orbit} shows one possible
configuration for an illustrative, planar 81-satellite
constellation----all placed in the orbital plane, at a mean cluster
altitude of 650 km. The arrangement here is based on a square rather
than hexagonal lattice, mostly to simplify its description. Cluster
radius is R=1 km, with distance between next-nearest-neighbor
satellites oscillating between (approximately) 100 and 200 m, as is
shown in Fig. \ref{fig:nndist8}.  We note that, of course, evolving
constraints could change the optimal architecture for our
constellation.

The constellation performs two shape-cycles per full orbit. Peripheral
satellite S1 is at apoapsis at $T=3T_{\text{orbit}}/12$, at altitude
$h=a+R/2$ and at periapsis at $T=9T_{\text{orbit}}/12$, at altitude
$h=a-R/2$. Since the diameter of S1's orbital ellipse equals that of
S0's, their orbital periods are identical; likewise for other
satellites. (Perturbations due to the J2 term and other effects will
require slight amendments to this leading-order-effect statement.)

While the constellation remains bounded inside a sphere of radius R,
its shape goes through two full cycles over the course of one
orbit. At any point in time, all satellites fit into a rotating
``$\pm\text{R}$ prograde, $\pm\text{R/2}$ in altitude'' ellipse. The
interior of this rotating ellipse does not perform a rigid rotation:
During the course of one orbit, different satellites are closest to
the endpoints of its semi-major/minor axes, as one readily observes by
following the path of S1.

If satellite motion were perfectly Keplerian (i.e. if Earth's
gravitational field were that of a point mass, and effects such as
solar and lunar tides, atmospheric friction, radiation pressure,
etc. were absent), this ``free fall'' constellation would reproduce
itself perfectly after a full orbit, at zero delta-v requirement. If
maintaining a planar constellation is undesirable, such as for
establishing inter-satellite links or to address passive safety
concerns, in-plane motion can be superimposed with per-satellite
oscillatory out-of-plane motion (one oscillation per orbit). The 2:1
axis ratio of the in-plane bounding ellipse is a consequence of
orbital dynamics. For a given minimal inter-satellite distance, this
design approach leads to the number of satellites scaling
quadratically with cluster radius, $N \sim R^2$. Improving the ratio
of sunlight capture area over cluster cross-section beyond quadratic
scaling would in general, for given minimal inter-satellite distance,
require non-Keplerian formation flight approaches---such as
electromagnetic formation flight \cite{Kong2004EMFF}---and would have
to pay attention to occlusion of outgoing ``rejected heat'' IR
radiation between satellites.

Taking the leading ``oblateness''-related J2-correction to Earth's
gravitational field into account, which here would be exploited to
keep satellites in a sun-synchronous orbit (i.e. make the orbital
plane rotate once per year), the cluster's shape would get deformed
slightly over the course of an orbit. This (predictable) drift can be
compensated for via a small adjustment to cluster shape. A simplistic
numerical calculation establishes that, for an example cluster as
described, adjusting the axis-ratio to 2:1.0037 can reduce J2-drift to
<3 m/s/year per km of maximal distance from reference orbit. A more
in-depth analysis of differential accelerations
(e.g. \cite{damico2010autonomous}) suggests that formation flight
should be feasible with only modest delta-v requirements beyond what
would be needed for precise station-keeping of a single satellite.

\subsection{TPU radiation testing}

Commercial-Off-The-Shelf (COTS) hardware has seen increasing use for
space missions \cite{sinclair2013radiation}, such as the Mars
Ingenuity helicopter \cite{Balaram2018}. However, high-performance ML
accelerators, which are characterized by a cutting-edge process node,
large die size, and high floating-point operations per second (FLOPS)
capability, represent a new frontier for COTS hardware in space. To
address the question of their viability, Google's V6e Trillium Cloud
TPU with its associated AMD host server were tested using a
\SI{67}{MeV} proton beam (with the beam energy at the device under
test being somewhat lower due to intervening materials; see Methods)
to simulate the operating conditions of sun-synchronous
LEO. This work presents the first published radiation-testing results
for such a device. For the target sun-synchronous LEO with significant
shielding (e.g. 10 mm Al equivalent), the radiation environment is
primarily composed of penetrating protons and Galactic Cosmic Rays
(GCRs) \cite{barth2003space}, resulting in an estimated dose of
$\sim$150 rad(Si)/year using industry-standard space radiation
analysis tools\cite{sinclair2013radiation}. Radiation mainly
causes: 1) Total Ionizing Dose (TID) effects, the cumulative build-up
of charge in insulating layers leading to device degradation
\cite{schwank2008radiation}, and 2) Single Event Effects (SEEs), which
are instantaneous faults caused by a single energetic particle strike
generating a dense track of electron-hole pairs
\cite{Xiong2023}. Sensitivity depends on the process node and cell
libraries used, and has historically decreased, but this trend does
not hold beyond the 5nm node \cite{Xiong2023}. For TPUs to be useful
for machine learning in space, they must survive the expected 5 year
lifetime TID of $\sim$750 rad(Si), and SEEs must be at levels
manageable through techniques like memory error correcting codes
(ECC). Undetected SEEs are a particular concern for model training, as
undetected bit-flips can lead to Silent Data Corruption (SDC). SDC may
occur in any operating environment; SEE-induced SDCs are not the only
cause \cite{dixit2021silent}. The covert nature of these errors
requires careful calibration, detection and correction techniques.

Of the different elements of the Trillium TPU, the High Bandwidth
Memory (HBM) subsystems exhibited the most sensitivity to TID. In our
tests, HBM-specific stress tests began to show irregularities after a
cumulative dose of 2 krad(Si), which is almost $3\times$ our minimum
requirement for TID survival of 750 rad(Si). All other tests,
including end-to-end ML workloads and compute-centric tests, continued
to operate correctly up to the maximum tested cumulative dose of 15
krad(Si) on a single chip. No hard failures were attributable to TID
up to this level.

Core logic and on-chip SRAM were the most SEE-sensitive components,
primarily manifesting as Silent Data Corruption (SDC) during end-to-end ML
workloads. For typical transformer workloads,
the observed SDC rate of approximately one event/17 rad, at an
estimated orbital dose of 150 rad(Si)/year, translates to a very low
failure probability, on the order of 1 per 3 million
inferences, assuming 1 inference per second. While this error rate is
likely acceptable for inference,
the impact of SEEs on training jobs, and the efficacy of system-level
mitigations, requires further study. Radiation testing of the host
system, with a focus on Single Event Functional Interrupts (SEFIs)
that manifest as system crashes or reboots, suggests a rate of one
such event per 450 rad(Si) for
CPU and 400 rad(Si) for RAM.

\subsection{Launch cost analysis}

High launch costs have historically stymied efforts to harvest solar
power in space on large scales \cite{Rodgers2024SpaceBased}. The below
does not constitute a full economic analysis, but gives proof points
demonstrating that there is a feasible path for launch costs to drop
sufficiently to no longer be prohibitive. Further,
\$200/kg is often cited, by SpaceX and others, as a threshold beyond
which launch could cease to be the limiting cost factor for ambitious
programmes
\cite{janson2024equatorial,COOPERSMITH201177,VirtusSolis2023Starship}.

We compared the kg/kW solar power (bus and payload) launched ratios
for Starlink v2, Starlink v1, OneWeb \& Iridium satellites and found
that, if LEO launch prices drop to \$200/kg, we can project \$/kW/y
launched to LEO (``launched power price'') could be $\sim\$810$/kW/y for
a Starlink v2-type constellation (amortized over satellite lifetime),
or $\sim$\$810--7,500/kW/y if we include a broader range of satellites
with different mass/power ratios to serve varied use cases. For
comparison, current power spend for terrestrial data centers in the US
is reported to be $\sim$\$570--3,000/kW/y (depending on regional
variation in power price and operator power usage effectiveness, or
PUE \cite{Nlyte2024DataCenter, shehab2025unitedstates,
  Google2024EnvironmentalImpact}). Thus, if launch costs to LEO reach
\$200/kg, then the cost of launch amortized over spacecraft lifetime
could be roughly comparable to data center energy costs, on a per-kW
basis.

\begin{figure}[htbp]
  \centering
\includegraphics[width=\columnwidth]{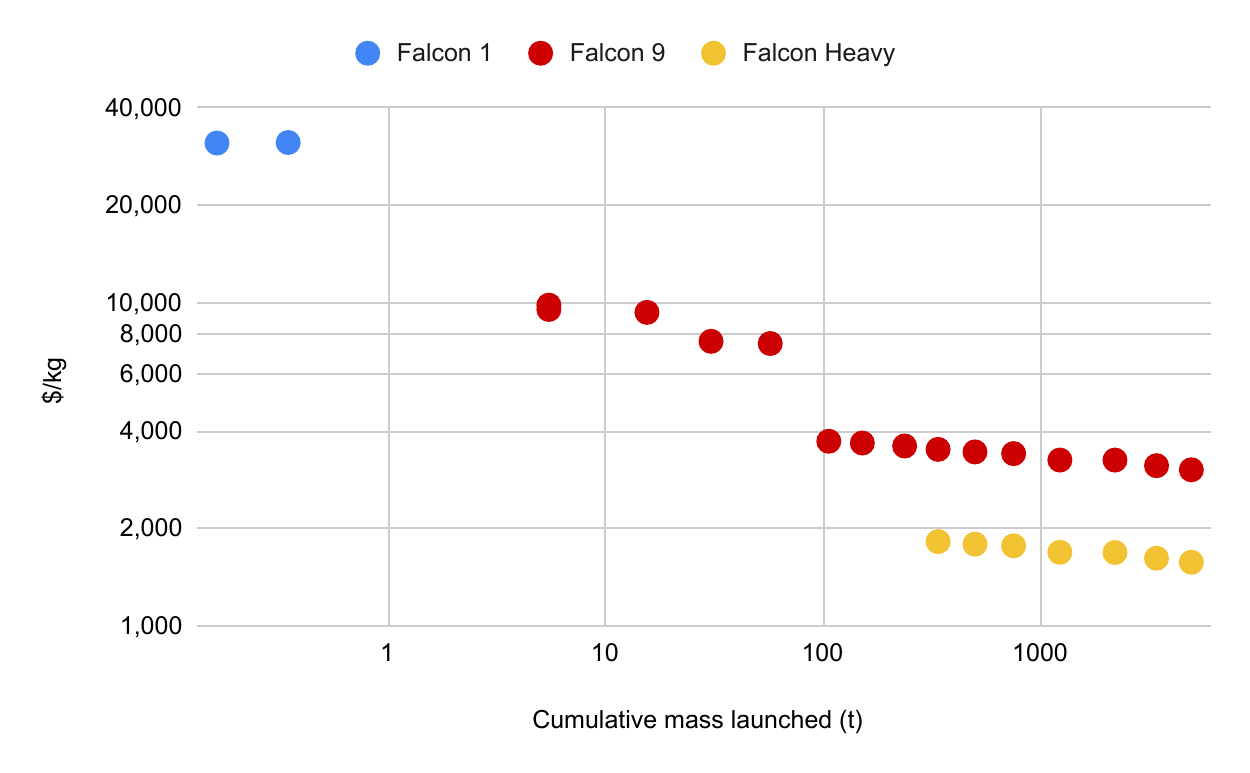}
  \caption{SpaceX payload mass launched by lowest achieved price, inflation-adjusted, since the first successful Falcon 1 launch, for progressive rocket categories. Note major price discontinuities at the introduction of Falcon 9 and Falcon Heavy \cite{NextSpaceflightLaunches, jones2018recent, JonathanSpaceReport}.}
  \label{fig:launched_mass}
\end{figure}

SpaceX launch pricing data and mass launched from Falcon 1 to Falcon
Heavy [Fig. \ref{fig:launched_mass}] yields a $\sim$20\% learning
rate, meaning the price per kg falls by $\sim$20\% for every doubling
of cumulative mass launched (over all vehicle classes). If the
learning rate is sustained---which would require $\sim$180 Starship
launches/year---launch prices could fall to <\$200/kg by
$\sim$2035 (taking the initial \$/kg price of Falcon Heavy as a starting point). While this would be a substantial achievement for SpaceX
(particularly given the technological discontinuity inherent in
switching to Starship), it is still far below stated launch targets
for Starship. Even if this launch rate is reduced by $\sim70\%$,
prices could drop to \$300/kg in the same timeframe, which would still
have a substantial impact on feasibility of large-scale
constellations. Further, there is precedent for a sustained $\sim$20\%
learning rate over multiple decades in other advanced industries
leveraging mass-production (notably, solar panels)
\cite{nijsse2023momentum}. Given the long lead times required to reach
scale for this type of ambitious project, it's strategically
beneficial to commence work on early milestones in anticipation of
projected price declines.

Alternatively, our analysis of Starship 4 public specifications and
data suggests that SpaceX launch costs to LEO may drop to
$\lesssim\$60$/kg ($10\times$ component reuse); if SpaceX's
$100\times$ component reuse target were achieved, costs could reach
$\lesssim\$15$/kg (assuming LOX prices of \$200/t and liquid methane prices $\lesssim\$700$/t, given SpaceX will use liquid methane for Starship instead of kerosene-based RP-1 at \$700/t \cite{Cunningham2023SpaceXFuel}, and liquid methane is less expensive than
kerosene when comparing it on a \$/megajoule basis \cite{daswani2022space}). Assuming $10\times$ reuse and even the highest
estimates of current SpaceX margins (up to 75\%
\cite{WRAL2024SpaceXCost}), launch price to customers would drop to
<\$250/kg (although margins are currently supported by SpaceX's
near-monopoly and hence likely to decrease with the anticipated entry
of competitors such as Blue Origin
\cite{techreview_spacex_rivals}). Realizing these projected launch
costs is of course dependent on SpaceX and other vendors achieving
high rates of reuse with large, cost-effective launch vehicles such as
Starship.

\section{Discussion}

The results presented here are a first milestone towards scalable
space-based AI; subsequent milestones involve testing aspects of the
system in space (other Google/Alphabet research initiatives, such as
Waymo and quantum computing, use similar milestone-based
approaches). These initial results are an encouraging first step in
assessing the feasibility of space-based ML compute at scale,
addressing fundamental challenges in inter-satellite communication,
orbital formation control, radiation hardness, and launch
economics. Future space-based experimental milestones should also
involve solutions for thermal management, high-bandwidth ground
communications, and on-orbit reliability and repair strategies. In
the supplementary material, we explore further work on orbital dynamics
modeling.

Effective thermal management is a critical optimization challenge for
power-dense TPUs operating in a vacuum. Advanced thermal interface
materials and heat transport mechanisms, preferably passive to
maximize reliability, are essential to efficiently move large heat
loads from the chips to dedicated radiator surfaces.

Managing potential failures and (relatedly) increasing reliability of
compute in space will be critical. Currently, failed TPUs are manually
replaced by technicians, which is relatively simple and low-cost on
Earth, but obviously impracticable in space. The simplest solution is
redundant provisioning. There are also promising research programs
aiming to increase system tolerance to faults affecting networking,
e.g. via reduced communication \cite{douillard2023diloco}.

Developing robust optical satellite-ground communications will also be
critical for scaled operation, but will necessitate overcoming
challenges including atmospheric turbulence, high-speed relative
motion errors, and precision beam tracking. This is an active field of
development \cite{CAILABS2024Keraunos, MIT2022LaserLink} with the
current leader, NASA's TeraByte Infrared Delivery (TBIRD) mission,
demonstrating 200Gbps ground-LEO communications in 2023
\cite{riesing2023operations}.

Looking further ahead, while our proposed constellation design is
naturally suited for growing clusters in space, unlocking the full
potential of compute in orbit will likely require new design
approaches for individual satellites. While we anticipate launch costs
continuing to decrease as the industry scales, the floor on fuel price
means that there will always be an incentive to minimize mass. Our
system design work to this point assumes a relatively conventional,
discrete compute payload, satellite bus, thermal radiator, and solar
panel design. However, as has been seen in other industries (such as
smartphones), massively-scaled production motivates highly integrated
designs (such as the system-on-chip, or SoC). Eventually, scaled
space-based computing would similarly involve an integrated compute,
radiator, and power design based on next-generation architectures,
such as computational substrates based on neural cellular automata
\cite{mordvintsev2020growing}. The TPU-based satellite cluster
described here and the preliminary results we describe are the
beginning of unlocking that potential. Realizing the full scope of
this ambitious vision will require sustained research, iterative
refinement of our design, and the achievement of several critical
future milestones.

\section{Methods}
\subsection{Orbital dynamics analysis}

In order to establish potential feasibility of a given formation
flight proposal, it is generally useful to start from some
analytically tractable simplifying model, such as the
Hill-Clohessy-Wiltshire equations \cite{clohessy1960terminal}, that
describe orbital motion of a satellite relative to a circular
reference orbit in a Keplerian approximation, to leading order in
relative positions and velocities - or more refined generalizations
such as the Tschauner-Hempel equations (for arbitrary eccentricity)
\cite{Tschauner1965Rendezvous} or Vinti theory \cite{Vinti1961Theory}
(taking the J2 "earth oblateness" contribution into account). If a
promising approach can be identified based on such analytic methods,
it makes sense to then explore the nature and magnitude of corrections
caused by further physical effects that are understood but more
difficult to model analytically, in order to see if they are
compatible with reasonable mission delta-v budgets. Here, the two main
approaches are perturbation theory and numerics. They can be used
independently, such as to validate one another's predictions, and
often also in conjunction. This leads to advanced numerical methods
for formation flight planning.

An overview over the relative importance of non-Keplerian
contributions to satellite acceleration can be found in textbooks such
as \cite{Montenbruck2013Satellite}. For formation flight, the most
important question is how various physical effects affect different
satellites in a close-proximity constellation differently. At the
envisioned altitude, the by far most important such effect is expected
due to the J2-term of the geopotential, and potentially differential
atmospheric drag. Effects such as lunar tides will be strongly
suppressed by very small factors such as $r_{\text{cluster}} /
d_{\text{moon}}$.

For this feasibility study, we determined promising-looking formation
flight patterns by starting from the basic Hill-Clohessy-Wiltshire
approximation, then mostly handling the adjustments needed to apply
them to a sun-synchronous cluster orbit perturbatively, and obtaining
results via numerical simulation. Care needs to be taken to retain
good control of numerical accuracy. With a model that only involves
accelerations with numerically benign behavior, computing orbits to
centimeter accuracy vs. orbital diameters of order-of-magnitude $10^7$
meters requires results to be correct to at least 9 decimal
digits. While the discrepancy between such numerical modeling and real
world physics may well be substantially larger than this accuracy
target, having good numerical accuracy for models with simplified
physics matters for objectives such as estimating the magnitude of
some physical effect via a fully-numerical rather than a hybrid
numerical-plus-perturbative approach.

Fast and performant (such as GPU-based) floating-point arithmetic
generally only supports up to \texttt{binary64} floating-point
representations, which give us just short of 16 decimal digits of
precision. This strongly suggests use of a high order ODE integration
scheme. For this exploration, we use the eighth-order Runge-Kutta
\texttt{DOP853} method provided by SciPy's
\texttt{scipy.integrate.solve\_ivp} function.

\subsection{Inter-satellite link analysis}

The feasibility of high-bandwidth, short-range ISLs was assessed
through link budget analysis. The achievable data rate of a
photon-limited optical link is directly proportional to the received
signal power, assuming a constant number of photons-per-bit (PPB)
required for a given modulation scheme. In the far-field, the received
power scales inversely with the square of the distance ($P_r \propto
1/d^2$) due to beam divergence, for a fixed transmitter power and
aperture size. This relationship, and the resulting impact on maximum
bandwidth for different modulation schemes, is plotted in Figure
\ref{fig:isl}.

A survey of currently available and announced commercial OISL
technologies reveals they are typically designed for link distances of
thousands of kilometers, offering maximum data rates ranging from
1\,Gbps to 100\,Gbps. For example, Starlink's system operates at
$\sim$100\,Gbps over distances up to $\sim$5400\,km, limited by line
of sight distances in LEO. These systems (shown in the lower right of
Figure \ref{fig:isl}) do not meet the multi-Terabit per second
requirements for tightly coupled ML clusters.

Our approach leverages the enhanced link budget at short distances to
employ multi-channel DWDM systems using high-spectral-efficiency
coherent transceivers. These are commercially available, e.g. 400G
transceivers using PM 16-QAM modulation. Deploying these on a 100GHz
ITU frequency grid, a single aperture could support 24 channels (half
of the C-band), yielding 9.6\,Tbps bidirectional bandwidth. A tighter
75\,GHz grid spacing could potentially support 12.8\,Tbps per
aperture. Such transceivers typically require a received power on the
order of -20\,dBm per channel \cite{FScom_2024}, totaling approximately
0.24\,mW for a 24-channel system. For the confocal case ($a = 5$\,cm, $d =
5$\,km), limiting beam wander to 10\% of the aperture radius
requires a pointing accuracy of $\sim$1.0\,$\mu$rad, well within the
demonstrated capability of commercial optical inter-satellite link
terminals that routinely operate at distances of thousands of
kilometers.

In the far field, the received signal power ($P_R$) is estimated using
the Friis transmission formula for free-space optics:
\begin{equation}
P_{R} = P_{T} \cdot G_T \cdot G_R \cdot \left(\frac{\lambda}{4\pi d}\right)^2 \cdot L_{\text{other}}
\end{equation}
Where $P_T$ is the transmitted power (assumed to be 5W from a
commercial EDFA), $G_t$ and $G_r$ are the transmitter and receiver
antenna gain (both 105.1dB, assuming $\sim$80\% aperture efficiency
for a 10\,cm diameter telescope), $\lambda$ is the wavelength
(\SI{1.55}{\micro\meter}), $d$ is the inter-satellite distance,
$L_{\text{other}}$ captures other losses (-3dB). The beam divergence
angle ($\theta$) for a diffraction-limited aperture at
\SI{1.55}{\micro\meter} is approximately $\theta \approx 1.22\lambda/D
\approx 1.22 \times (1.55 \times 10^{-6}\,\text{m}) / 0.1\,\text{m}
\approx 18.9\,\mu\text{rad}$. To illustrate the power constraints of
existing long-range systems, consider a typical LEO-LEO link distance
of 5,000\,km. At this distance, the beam spot radius is at least 95
meters, and the best case received power \SI{1.6}{\micro\W}.

For the short-range, high-bandwidth links central to our spatially
multiplexed design, a near-field model provides a useful
approximation. In a symmetric confocal system where the beam waist is
located midway between the two transceivers, the link distance ($L$)
for a given beam radius ($a$) at the optics is given by: $L = \pi a^2
/ \lambda$. For example, for a 10\,cm diameter beam at the optics ($a =
5$\,cm), and a wavelength of \SI{1.55}{\micro\meter}, the link
distance for this system is approximately 5\,km under best-case
conditions (ignoring beam quality, aperture truncation and pointing
jitter).  This near-field analysis also shows that at kilometer-scale
distances, minimum required aperture size for a single link scales
with the square root of the distance. Consequently, more independent
links can be packaged into a fixed total area as the distance
decreases, causing the total bandwidth available via spatial
multiplexing to scale inversely with distance.

The PPB requirements for different schemes shown as lines in Figure
\ref{fig:isl}, illustrating the trade-off between distance and maximum
achievable bandwidth. They were estimated as $\sim$71 PPB for
On-Off Keying (OOK)\cite{laube2025apd}, and $\sim$196 PPB for
PM 16-QAM\cite{FScom_2024}, as this coherent scheme
requires a higher signal-to-noise ratio. This compares to the
Shannon-Hartley limit of $\sim$1.39 PPB, calculated in the infinite
bandwidth, shot-noise limit as $2 \times \ln(2)$). We assumed the use
of standard C-band and L-band wavelengths, consistent with mature DWDM
technology.

\subsection{TPU Radiation Testing Procedure}
Facility and Beam: Testing was conducted at the UC Davis Crocker
Nuclear Laboratory, utilizing their 76-inch cyclotron to produce a
\SI{67}{MeV} proton beam. A large 8\,cm diameter aperture was used to
ensure uniform irradiation of the entire TPU package, including the
logic die and HBM stacks. Beam intensities ranged from 2\,pA
($\sim$2\,rad/min) to 1\,nA (1\,krad/min).

Test Rig and Beam Path: A standard TPU system was tested with
heatsinks installed to permit full-power operation. To avoid beam
obstruction by the primary topside heatsink, chips were irradiated
from the underside. The proton beam traversed the chassis ($\sim$1mm
Al), a secondary heatsink assembly, and the printed circuit board
(PCB) before reaching the silicon die, which is directly bonded to the
PCB. The secondary heatsink was modified to ensure uniform irradiation
of the chip.

Beam Energy Considerations: The range of \SI{67}{MeV} protons,
calculated using the NIST PSTAR database \cite{Berger2017STAR}, is
approximately 18\,mm in aluminum and 6.5\,mm in copper. Therefore,
while most of the proton flux will reach the die (confirmed by imaging
through the complete assembly with gafchromic film
\cite{sorriaux2013evaluation}), the energy spectrum incident is not
monoenergetic \SI{67}{MeV} but is lower and spread out due to energy
loss and straggling within the heatsink material. Additionally,
nuclear interactions within the heatsink materials will generate
secondary particles (neutrons, gammas, etc.), contributing to the
radiation environment experienced by the device under test.

Software Suite: Subsystems were stressed with a diverse software suite
to assess different vulnerabilities. This included memory stress tests
targeting the High Bandwidth Memory (HBM) with various read/write
patterns, compute-centric tests focusing on the core tensor
computation units with minimal HBM traffic, and an end-to-end ML
workload running a transformer model to simulate realistic operational
stress. The memory tests were configured to log errors and continue,
while the compute-centric and end-to-end tests were set to halt on the
first significant error to establish dose-to-failure thresholds.

Data analysis: To estimate proton-induced SEE cross-sections, we used
the beam characteristics provided by the facility and dose
measurements. A dose of 1\,rad corresponds to a proton fluence of
approximately $7.9 \times 10^6$ protons/cm$^2$. The per-chip
cross-section ($\sigma$) was calculated as: $\sigma \approx 1.27
\times 10^{-7} / D \text{ cm}^2/\text{chip}$ where D is the dose per
event in rad.

Core logic and on-chip SRAM were the most SEE-sensitive components,
primarily manifesting as Silent Data Corruption (SDC) during end-to-end
test execution. The characteristic dose for SDC averaged between 14.4 rad
and 20 rad per event depending on the specific workload. This corresponds
to a cross-section of approximately $6 \times 10^{-9}$  to
$9 \times 10^{-9} cm^2$ / chip.

HBM errors manifested as
uncorrectable ECC errors (UECCs), causing test halts. The
characteristic dose for HBM UECC was approximately 44\,rad per event
(averaged over 203 events). This corresponds to a cross-section of
approximately $3 \times 10^{-9}\text{cm}^2 /\text{chip}$.
System-level crashes, indicative of Single Event
Functional Interrupts (SEFIs), were observed on average once per
5\,krad of dose per chip. This translates to a SEFI cross-section of
approximately $2 \times 10^{-11}\text{cm}^2/\text{chip}$.

It is important to note that HBM correctable error (CECC) counts were
not reliably available for this analysis, as single-bit CECCs are only
reported by the HBM firmware above a non-configurable, vendor-specific
threshold. Thus, a precise in-test CECC to UECC ratio for HBM could
not be determined. However, instances of data mismatches occurring
without corresponding UECC flags solidly demonstrate that SDCs originating
in the logic and SRAM dominate the observed soft error rate for the
System-on-Chip (SOC).

\subsection{Launch cost analysis}
\FloatBarrier
\begin{table*}[hbt!]
  \centering
  \caption{Launched power prices for a range of LEO satellites}
  \label{tab:launchcosts}
  \begin{tabular}{@{}lccccc@{}}
Satellite & \begin{tabular}[c]{@{}c@{}}Mass\\ (kg)\end{tabular} & \begin{tabular}[c]{@{}c@{}}Power\\ (kW)\end{tabular} & \begin{tabular}[c]{@{}c@{}}Lifespan\\ (y)\end{tabular} & \begin{tabular}[c]{@{}c@{}}Launched power \\ at \$3,600/kg \\ (\$/kW/y)\end{tabular} & \begin{tabular}[c]{@{}c@{}}Launched power \\ at \$200/kg  \\ (\$/kW/y)\end{tabular} \\ \midrule
    Starlink v2 mini [opt.] & 575 & 28 [est.] & 5 & \$14,700 & \$810 \\
    Starlink v1 & 260 & 7 [est.] & 5 & \$26,600 & \$1,470 \\
    OneWeb & 150 \cite{EoportalOneWeb} & 0.8 \cite{SatNews2025AirbusRocketLab} & 5 & \$135,800 & \$7,500 \\
    Iridium & 860 \cite{EoportalIridiumNEXT} & 2 \cite{SpaceNews2010IridiumSelects} & 12.5 & \$124,600 & \$6,900 \\
    \bottomrule
  \end{tabular}
\end{table*}

We projected launch prices via two methods: learning curve projection
and analysis of planned Starship 4 specifications and reuse
targets. While there is inherent uncertainty (e.g. in future
regulatory obstruction, competitive dynamics from new entrants into
the launch market and unforeseen technical challenges), both methods
support our conclusion that reaching customer prices of
$\lesssim$\$200/kg by mid 2030s is plausible under reasonable
assumptions for reuse and cumulative mass launched during the time. We
stress that the following is not intended as a comprehensive economic
feasibility study, but rather a high-level evaluation of the potential
for launch costs, specifically, to affect the viability of our
proposal.

Sustaining high learning rates over time is obviously challenging, but
there are multiple precedents across other advanced industries,
e.g. shipbuilding and aerospace \cite{brown2019learning}. In addition,
solar panels provide a particularly striking, canonical example of a
sustained $\sim$20\% learning rate for over 40 years
\cite{nijsse2023momentum}.

Sustaining high growth in launched mass rates (and attendant price
decline) is unlikely without large-scale, non-SpaceX commercial
constellations driving demand \cite{gatti2025missing,
  McKinsey2023SpaceLaunch}. Scaling ML in space would be one such use
case, providing consistent demand for Starship (or Starship-class)
launches.

Our learning curve analysis was based on publicly available historical
SpaceX data. When Falcon 1 launched, SpaceX launch prices were
$\gtrsim$\$30,000/kg (inflation adjusted)
\cite{SpaceNews2013SmallLauncher}, dropping to $\sim$\$1800/kg for
Falcon Heavy over the course of <100 successful launches
\cite{Chang2018FalconHeavy}, or $\sim$400t cumulative mass launched
[Fig. \ref{fig:launched_mass}]. The introduction of Falcon Heavy
yielded a precipitous price decrease compared to Falcon 9, driven
by improved economics from a heavier launch. Note that learning
curve estimates are highly sensitive to input assumptions
(e.g. source for price data, or whether the chosen starting point
is the first successful Falcon 1 launch, the Falcon 9 or even the
reusable Falcon 9 configuration), but a variety of choices all
yield results $\sim$18--24\%.

Maintaining SpaceX's $\sim$20\% learning rate (based on launched mass)
through new generations of launch vehicles, reaching <\$200/kg by
$\sim$2035 would require launching $\sim370,000\text{t}$ additional
cumulative mass, equivalent to $\sim$1800 Starship launches (assuming
200t capacity (original specs at \cite{SpaceX_2024}, updated
\cite{Musk2025LaunchCosts}). This is an ambitious target, but the
required $\sim$180 Starship launches/year (on average, although there
will almost certainly be a ramp up period in reality) would fall well
below stated launch rate targets \cite{SpaceXMars}. Conversely,
reaching this target is unlikely without Starship-like launch
vehicles, due to Falcon payload volume limitations. We also stress
that the calculation of required launched mass to reach our price target is
highly sensitive to chosen initial conditions. We have selected the introduction
of Falcon Heavy (i.e. $\sim$\$1800/kg and $\sim$400t cumulative mass launched)
as our starting point, on the basis that, while all price trajectories are
inherently highly uncertain and subject to market forces (such as the entry -or
not- of competitors), price at the introduction of a new launch vehicle represents
a critical inflection point and is likely tied to some underlying economic reality.
In the above analysis, we are not trying to predict the actual future price of launch,
but rather make a reasonable assessment of what the underlying economics could theoretically support.

As an alternative method, we calculated costs (to SpaceX) for Starship
4, leveraging recently released reuse targets, as well as plans for
payload size, Raptor engine count and rocket dimensions
\cite{Musk2025LaunchCosts}. Other inputs, e.g. refurbishment costs as
a fraction of vehicle costs ($\sim$1\%
\cite{Marketplace2024CheaperSpaceFlight}) and failure rates, were
extrapolated from existing Falcon 9 data \cite{WRAL2024SpaceXCost,
  Marketplace2024CheaperSpaceFlight,Musk2018Falcon9}. Without assuming
any reduction in fuel cost, component reuse drives down projected
SpaceX launch costs from $\sim$\$460/kg (no reuse) to
<\textasciitilde\$15/kg (100x reuse of all components). Increasing
refurbishment costs to 15\%, as a sensitivity analysis, yields costs
of \$38/kg (100x reuse of all components). Eventually, fuel costs (LOX
and liquid methane for Starship), estimated at $\sim$\$8/kg
\cite{Cunningham2023SpaceXFuel, DLA2024AerospacePrices}, will provide
a floor on costs. We note that this analysis was based on relatively
conservative assumptions, e.g. we used published Raptor 3 engine
specifications, not any speculative claims for future engine
generations. We also acknowledge the potential pitfalls of using
Falcon 9 data to project trends for Starship, but the intent is to
provide an indicative projection based on public data points, which we
will update as more information becomes available.

The primary ingredients to run compute are power, infrastructure and
chips. To compare annualized cost per unit of power for space and
terrestrial data centers, we examine the launch costs for several
satellite types, in terms of \$/kW launched to LEO (``launched power
price''). A strict like-for-like comparison of satellite and
terrestrial costs would require more precise costs for satellite
design, but we can make an approximate comparison between data center
power costs and the launched power price. We do not include
infrastructure or building costs, since this is distinct from power,
or chip costs, since these will arise for both space and terrestrial
configurations.

For the following analysis, we use current launch price \$3,600/kg,
based on Falcon 9 (reusable configuration), since that is used for the
Starlink constellation, \cite{SpaceXFalcon9, Pelham2025SpaceXCost} and
``potential'' price \$200/kg. Note that we based this analysis purely on
SpaceX data since they are by far the leading launch provider. Other
incumbent and new entrants into the market (e.g. RocketLab and Blue
Origin) are likely only to hasten price declines via competition and
erosion of high SpaceX margins \cite{WRAL2024SpaceXCost}.

Starlink is the best-in-class proxy, as the largest constellation in
orbit. The new v2 mini satellites weigh 575kg (optimized design
\cite{Starlink2023Progress}). Exact specifications have not been
publicly released by SpaceX, but photometric and other analyses yield
solar panel area $\sim$105m$^2$ \cite{mallama2023starlink,
  Rodgers2024SpaceBased}. Assuming 22\% solar panel efficiency,
1.361kW/m$^2$ solar insolation and 90\% packing area for square cells
\cite{johnson2015lightweight, Sinovoltaics2025PackingDensity,
  Elkelawy2025PVPlant}, we obtain $\sim$28kW/satellite. Amortized over
a 5 year lifespan \cite{Wall2023StarlinkSatellites}, we obtain a
launched power price of \$14,700/kW/y (current prices), falling to
\$810/kW/y if launch drops to \$200/kg. We performed similar
calculations for Starlink 1, OneWeb, and Iridium NEXT satellites
(see [Table \ref{tab:launchcosts}]), yielding a launched power price
range (for launch prices \$200/kg to LEO) of \$810--7,500/kW/y. Note
that this extremely large range is at least partially driven by
differences in use case and resulting optimization priorities across
satellite programs, which in turn affects design decisions and kg/kW
ratio.

By comparison, terrestrial power costs for ML-capable data centers in
the US are reported to be \$0.06--0.25/kWh \cite{Nlyte2024DataCenter}
and PUE ranges $\sim$1.09--1.4 \cite{shehab2025unitedstates,
  Google2024EnvironmentalImpact}, which yields annual power spend of
\$570--3,000/kW/y. That is, if launch costs reach $\lesssim$\$200/kg,
annualized cost per unit of power in space could be approximately
comparable to terrestrial spend. If, instead, 104,000t are launched (a
72\% decrease from the above target), prices could be $\sim$\$300/kg
by the mid-2030s. The launched power price for a constellation of
Starlink v2 mini satellites would then be $\sim$\$1,200/kW/y, still
within range of terrestrial annual power prices.

\clearpage

\bibliography{Suncatcher}

\section*{Acknowledgements}
We thank Amaan Pirani for critical contributions to cost
modeling and overall feasibility analysis, Marcin Kowalczyk for
independent numerical validation calculations, Paul Epp and Stephen
Palese for input on the ISL concept, Thomas Zurbuchen for his
contributions to the systems and architecture concepts, and Kenny
Vassigh and Jerry Chiu for technical input on system and thermal
design. We also thank Muon Space for general discussions and for
technical and economic feasibility analysis of the concept.

\section*{Author Information}
These authors contributed equally: B.A.A., T.R.B., M.B., J.V.B., T.F., K.G., U.K., R.P.

B.A.A. conceived of this overall project.
T.F. developed the orbital dynamics and formation flight work.
U.K. and R.P. designed and performed the radiation testing.
J.V.B. conducted the launch cost
analysis. J.V.B., U.K. and T.R.B. prepared and proofread the
manuscript. K.G. and U.K. conducted the inter-satellite link
analysis. T.R.B. and M.B. developed the system design
overview. J.M. provided overall guidance and supervision to the
project.

\clearpage
\section*{Supplementary Information}
\subsection*{Orbital dynamics modeling}

Upper bounds on mission requirements (such as delta-v requirements)
can often be derived from conservative estimates based on simplifying
models. When implementing a control model for satellite cluster
formation flight, accurate information about satellite position and
orientation, precise models of predictable accelerations (due to
Earth's known but irregular gravitational potential, solar and lunar
tides, etc.), and reasonable models of noisy accelerations (such as
due to space weather- dependent atmospheric friction) are
indispensable: every in-principle-manageable acceleration that is not
included in the control model will manifest as unexpected drift,
increasing overall mission delta-v requirements.

One promising approach towards implementing formation flight control
is to use backpropagation-based techniques. This in general starts
from an objective function whose calculation involves numerical
ODE-integration that utilizes the complete motion-state of all
satellites in the formation, and whose minimum describes a desirable
target-state. This function will in general accumulate (transient)
violations of the cluster being in a good configuration. If the
control---whose output can be understood as a plan to drive actuators---is
implemented in terms of an algorithm with tunable parameters (and may
include a learned model), adjoint-state methods can be used to
backpropagate objective-function gradients through ODE-integration,
and automatic differentiation (typically reverse-mode AD) can then
backpropagate gradients into model parameters. Implementing such an
approach is greatly simplified by employing a Machine Learning
framework such as JAX. This approach can also handle higher-order
derivatives via ``differentiable programming'' approaches.

\end{document}